\documentclass[conference]{IEEEtran}
\IEEEoverridecommandlockouts
\usepackage{cite}
\usepackage{amsmath,amssymb,amsfonts}
\usepackage{algorithmicx}
\usepackage{graphicx}
\usepackage{textcomp}
\usepackage{xcolor}
\usepackage{multirow}
\def\BibTeX{{\rm B\kern-.05em{\sc i\kern-.025em b}\kern-.08em
    T\kern-.1667em\lower.7ex\hbox{E}\kern-.125emX}}
    
\usepackage[utf8]{inputenc} 
\usepackage[T1]{fontenc}    
\usepackage{hyperref}       
\usepackage{url}            
\usepackage{booktabs}       
\usepackage{amsfonts}       
\usepackage{nicefrac}       
\usepackage{microtype}      
\usepackage{lipsum}		
\usepackage{subcaption}
\usepackage[square,sort,comma,numbers]{natbib}
\usepackage{doi}
\usepackage{gensymb}
\usepackage{algpseudocode}
\usepackage{amsmath}
\usepackage{color,soul}
\usepackage{siunitx}
\usepackage{enumitem}
\usepackage{multirow}
\newcommand{\vv}[1]{\textbf{#1}}

\newcommand{\PretrainingEpochs}{80\,}
\newcommand{\ScalingEpochs}{5\,}

\newcommand{\PeakFLOPJUWELSScale}{3,072\,}
\newcommand{\PeakFLOPJUWELS}{99.1\,}
\newcommand{\PeakFLOPPerlmutterScale}{3,808\,}
\newcommand{\PeakFLOPPerlmutter}{140.8\,}
\newcommand{\PeakFLOPPerlmutterEfficiency}{11.9\%\,}
\newcommand{\PeakFLOPSeleneScale}{1,024\,}
\newcommand{\PeakFLOPSelene}{91.8\,}

\newcommand{\EnsInferenceTime}{12.4\,}
\newcommand{\InferenceSpeedup}{80,000\,}
\newcommand{\InferenceSpeedupJ}{80,000X\,}
\newcommand{\TimeToSolutionJUWELSTrainingScale}{3,072\,}
\newcommand{\TimeToSolutionJUWELSTrainingMinutes}{67.4\,}

\makeatletter
\newcommand{\customlabel}[2]{%
   \protected@write \@auxout {}{\string \newlabel {#1}{{#2}{\thepage}{#2}{#1}{}} }%
   \hypertarget{#1}{}
}
\makeatother

\begin{document}

\title{FourCastNet: Accelerating Global High-Resolution Weather Forecasting using Adaptive Fourier Neural Operators}

\author{\IEEEauthorblockN{Thorsten Kurth}
\IEEEauthorblockA{\textit{}
\textit{NVIDIA}\\
Zurich, Switzerland \\
tkurth@nvidia.com}
\and
\IEEEauthorblockN{Shashank Subramanian \textsuperscript{$\star$}}
\IEEEauthorblockA{\textit{}
\textit{NERSC, Lawrence Berkeley Lab}\\
Berkeley, CA, USA \\
shashanksubramanian@lbl.gov}
\and
\IEEEauthorblockN{Peter Harrington \textsuperscript{$\star$}}
\IEEEauthorblockA{\textit{}
\textit{NERSC, Lawrence Berkeley Lab}\\
Berkeley, CA, USA \\
pharrington@lbl.gov}
\and
\IEEEauthorblockN{Jaideep Pathak \textsuperscript{$\star$}}
\IEEEauthorblockA{\textit{}
\textit{NVIDIA}\\
Santa Clara, CA, USA \\
jpathak@nvidia.com}
\and
\IEEEauthorblockN{Morteza Mardani}
\IEEEauthorblockA{\textit{}
\textit{NVIDIA}\\
Santa Clara, CA, USA \\
mmardani@nvidia.com}
\and
\IEEEauthorblockN{David Hall}
\IEEEauthorblockA{\textit{}
\textit{NVIDIA}\\
Santa Clara, CA, USA \\
dhall@nvidia.com}
\and
\IEEEauthorblockN{Andrea Miele}
\IEEEauthorblockA{\textit{}
\textit{NVIDIA}\\
Santa Clara, CA, USA \\
amiele@nvidia.com}
\and
\IEEEauthorblockN{Karthik Kashinath}
\IEEEauthorblockA{\textit{}
\textit{NVIDIA}\\
Santa Clara, CA, USA \\
kkashinath@nvidia.com}
\and
\IEEEauthorblockN{Animashree Anandkumar}
\IEEEauthorblockA{\textit{}
\textit{NVIDIA and Caltech}\\
Santa Clara, CA, USA \\
aanandkumar@nvidia.com}
}

\maketitle
\begingroup\renewcommand\thefootnote{$\star$}
\footnotetext{Equal contribution}

\begin{abstract}
Extreme weather amplified by climate change is causing increasingly devastating impacts across the globe. The current use of physics-based numerical weather prediction (NWP) limits accuracy due to high computational cost and strict time-to-solution limits.

We report that a data-driven deep learning Earth system emulator, FourCastNet, can predict global weather and generate medium-range forecasts five orders-of-magnitude faster than NWP while approaching state-of-the-art accuracy. FourCastNet is optimized and scales efficiently on three supercomputing systems: Selene, Perlmutter, and JUWELS Booster up to \PeakFLOPPerlmutterScale NVIDIA A100 GPUs, attaining \PeakFLOPPerlmutter petaFLOPS in mixed precision (\PeakFLOPPerlmutterEfficiency of peak at that scale). The time-to-solution for training FourCastNet measured on JUWELS Booster on \TimeToSolutionJUWELSTrainingScale GPUs is \TimeToSolutionJUWELSTrainingMinutes minutes, resulting in an \InferenceSpeedup times faster time-to-solution relative to state-of-the-art NWP, in inference. 

FourCastNet produces accurate instantaneous weather predictions for a week in advance, enables enormous ensembles that better capture weather extremes, and supports higher global forecast resolutions. 
\end{abstract}


\section{Highlights}
\label{sec:highlights}
The project measures global weather emulation by time-to-solution. Training time on \TimeToSolutionJUWELSTrainingScale GPUs is \TimeToSolutionJUWELSTrainingMinutes minutes and scalability demonstrated to \PeakFLOPPerlmutterScale GPUs at 8X higher resolution than state-of-the-art deep learning models. 24-hour 100-member ensemble forecast time reduced from 984,000 to \EnsInferenceTime node-seconds, accelerating \InferenceSpeedupJ versus state-of-the-art numerical simulations.

\section{Overview}
\label{sec:overview}

Climate change is amplifying extreme weather events by increasing their frequency and intensity and changing their spatio-temporal and geographical patterns resulting in unprecedented stress on natural and human ecosystems. Munich RE estimates nearly a million human deaths and US\$4.2 trillion in damages due to extreme weather events since 1980 and projects damages in 2050 to rise to US\$1.7 trillion \emph{per year}. Swiss RE forecasts up to 18\% loss in global economic output by 2050 if no climate change-mitigating actions are taken.

Preserving both human and ecosystem well-being depends on our ability to predict, mitigate, and adapt to climate change. Weather forecasts and climate predictions help plan critical sectors such as transportation, logistics, agriculture, energy, and life-saving humanitarian response. 

For several decades, scientists have used numerical prediction models that simulate the dynamics of multiple interacting components of the Earth system, including the atmosphere, ocean, sea ice, land ice, and vegetation. The accuracy of these models has been steadily improving due to increasing resolution, better characterization of physical processes, and improvements in atmospheric observations leading to better model initialization via data assimilation \citep{bauer2015quiet}. 

Today, weather and climate models are extreme-scale supercomputing applications consisting of millions of lines of code which demand exascale computing. In order to improve the precision and accuracy of these critical models we must contend with eight grand challenges:

1. \emph{Complexity}: The Earth system exhibits multiple complex physical processes including turbulence, radiation, multi-phase multi-scale multi-physics, chemistry, and biology. Accurately characterizing all such processes in a numerical model would require the solution of hundreds of coupled nonlinear partial differential equations (PDE), which is prohibitively expensive. 

2. \emph{Resolution}: Interactions across scales in the Earth system determine cascades of energy, momentum, and entropy that govern the turbulent dynamics of weather systems. These interactions span more than twelve orders of magnitude in space and time, from micrometers at the molecular scale to thousands of kilometers at the planetary scale. The resolution of a weather or climate model determines which scales and physical processes can be explicitly resolved and which must be approximated. Model fidelity increases with resolution. Resolving storms at scales that matter for societal impacts requires kilometer-scale or finer \cite{neumann2019assessing}. Computational demands of weather and climate models, however, scale with the fourth power of resolution. Furthermore, the greatest source of uncertainty in climate projections has been attributed to uncertainties in shallow cumulus and stratocumulus clouds. The resolution required to explicitly represent these cloud processes is on the order of 1 m, which would require 100 billion times more computational resources than today's 1 km weather and climate models. Based on the continued exponential growth of computing resources, achieving 1 m global resolution will not be possible before the year 2060 \cite{Schneider2017}.

3. \emph{Dimensionality}: Numerical weather and climate models have very high dimensionality; at 1 km resolution the trajectories of complex nonlinear weather and climate models traverse a tera-dimensional phase space ($10^{12}$ degrees of freedom). This high dimensionality produces enormous state vectors which are extremely difficult to move, store, and analyze.

4. \emph{Ensemble size}: Large ensembles of simulations are required to adequately characterize the distribution of possible outcomes, due to initial and boundary condition uncertainties, model or structural uncertainties, and uncertainties produced by the stochasticity of fine-scale physical processes. Furthermore, inherently rare and extreme events that lie at the tails of the distribution can be confidently predicted only with well-calibrated large ensembles, further amplifying the computational cost of numerical simulations. 

5. \emph{Scenario diversity}: The many possible pathways of human behavior, including potential mitigation strategies, greatly expands the number of scenarios that need to be considered to produce accurate climate projections. Nonlinear feedback between hundreds of tightly coupled processes mean that different patterns of human behavior can result in vastly different outcomes. Accurately exploring this dimension requires the simulation of orders-of-magnitude more scenarios than is currently possible.  

6. \emph{Throughput}: The workflow of a numerical weather forecast includes processing hundreds of observational datasets, data assimilation to create accurate initial conditions for forecasts, time integration of the numerical weather forecast model, as well as post-processing and dissemination of forecast products. The demands for multiple operational weather forecasts several times a day amounts to at least one simulated year per day (SYPD) using thousands of nodes \cite{neumann2019assessing}. The workflow of a climate model involves long-running simulations, data storage, and climate diagnostics and analyses. The resolution, set of scenarios and experiments, ensemble sizes, and multi-decadal timescales of climate projections also translate to 1 SYPD using thousands of nodes \cite{neumann2019assessing}. 
In order to achieve this throughput at 1 m resolution for the requirements listed above, however, weather and climate models need to be on the order of 100 billion times faster \cite{Schneider2017}. 

7. \emph{Scalability and performance}: Kilometer-scale simulations are now feasible, given the advances in high-performance computing (HPC). Yet, only a handful of studies of global kilometer-scale simulations and their computational performance are available due to fundamental, unresolved challenges in the scalability and performance of weather and climate models on current leadership class HPC systems \cite{Wedi2020}. Furthermore, weather and climate models are neither prepared for the transition to supercomputers designed and optimized for AI workloads at single/mixed precision, nor are they capable of efficiently leveraging GPU-acceleration on leadership class HPC systems. Simultaneously, the energy consumption of weather and climate simulations reveals a 100-to-1 ratio between data transfers and compute, thus data movement has become the dominant performance constraint \cite{Schar2020}. This bottleneck is further exacerbated as resolution increases. 

8. \emph{Flexibility and interactivity}: The Earth system modeling enterprise, including weather forecasting and climate predictions, generates tens of petabytes of data every year and that quantity is growing exponentially. Yet, the gigantic volume of data and wealth of information contained therein is often collapsed into a few simple metrics that guide decision-making. The community of experts, scientists, researchers, and policy-makers desperately need flexible and interactive tools that can unfold and extract information from this high-dimensional multi-variate data space in order to make better informed decisions. Digital twins are virtual replicas of real-world systems that enable interactivity and visualization of potential futures through intuitive and responsive user interfaces \cite{Bauer2021}. Realizing a complete digital twin of the Earth, however, will require unprecedented advancements in all the grand challenges listed above. 

We frame the problem statement as: \textit{How can the eight grand challenges be addressed in a unified way using alternatives to traditional numerical Earth system modeling in order to help realize ``Interactivity at Scale'' via digital twin Earth?}

The first step is achieving a million times greater throughput in global high-resolution weather forecasting and climate prediction, a problem that state-of-the-art deep learning (DL) is well-poised to address. 

\section{State Of The Art}
\label{sec:state_of_the_art}

Two areas influence the current landscape of weather and climate modeling: numerical prediction and deep learning.

\subsection{Numerical Prediction}
Advances in HPC have allowed a doubling in the performance of weather and climate simulations every two years \cite{Schulthess2019}, enabling today's global weather and climate simulations to be run at kilometer scale. 

Wedi et al. \cite{Wedi2020} report accelerating four-month long IFS simulations on Summit and on Piz Daint at 1.4 km resolution, 62-vertical levels, both hydrostatic and nonhydrostatic versions, using CPUs only with a hybrid MPI/OpenMP parallelization. The scalability progress in the IFS combined with Summit's computing potential helped them achieve an impressive throughput of 0.3 SYPD for the hydrostatic version on close to 4,000 nodes on Summit and 0.085 SYPD for the non-hydrostatic version on close to 5,000 nodes on Piz Daint (cf. Figure 1 in \cite{Wedi2020}). They anticipate that with the use of GPUs the throughput will increase even further, though ensembles at 1.4 km grid spacing will face substantial computational constraints.

Neumann et al. \cite{neumann2019assessing} report on the performance of ICON global simulations at kilometer scale with the caveat that the exact complexity of model components such as land surface, sea ice, or atmospheric chemistry; the exact vertical resolution; or the time-step length are not yet settled. Due to memory requirements, 1 km global simulations can only be run on a minimum of approximately 1,000 nodes. For reasons described in \cite{neumann2019assessing}, primarily due to load imbalances, scalability of the 1 km global climate simulation stagnates at O(200,000) nodes at a throughput of approximately 0.06 SYPD, extrapolated from the 5 km case using a performance model that assumes perfect weak scaling and neglects I/O.

The state-of-the-art for single runs of numerical weather and climate simulations is approaching 1 SYPD on thousands of nodes of leadership class supercomputing systems. The associated energy costs of global kilometer-scale simulations are estimated to be 596 MWh/SY for weather and 191.7 MWh/SY for climate \cite{Fuhrer2018, Schulthess2019}. Yet, thousands of runs are required to generate large ensembles, simulate different scenarios, and explore the impact of mitigation strategies, and the energy requirements are thousands of times larger as well.  

\subsection{Deep Learning}

\subsubsection{Operator learning and transformers}
Neural operators learn mappings between infinite dimensional spaces and have demonstrated impressive results in creating surrogate models of classical problems in fluid dynamics governed by the Navier-Stokes equations \cite{kovachki2021neural}. The Fourier Neural Operator (FNO) is a class of integral operators that efficiently implements global convolution via Fast Fourier Transform (FFT) which has been used very successfully to solve nonlinear and chaotic PDEs \cite{li2021physics}. Transformer models, the state-of-the-art in DL, are proving to be tremendously successful across diverse applications \cite{dosovitskiy2020image, khan2021transformers}. Unhindered by prescribed inductive biases, they naturally learn inherent inductive biases in large quantities of training data for complex data-driven tasks. In contrast with traditional convolutional neural networks (CNNs), transformer performance improves as the data and model sizes grow. Furthermore, the multi-head self-attention layers enable vision transformers (ViT) to model long-range dependencies and non-local feature dependencies with ease. These advantages make ViTs well-adapted for challenging tasks like super-resolution, image de-noising and de-blurring \cite{khan2021transformers}.

ViTs treat images as a sequence of small patches (tokens) and therefore rely on a key-value self-attention mechanism, which scales quadratically with the input image resolution \cite{vaswani2017attention}. When ViTs are applied to weather and climate data at high-spatial resolution (for example, 20,000$\times$40,000 pixels for 1 km resolution), the number of computations and amount of memory required becomes prohibitive. Recent research in DL and computer vision has attempted to remedy this inefficiency by approximating the self-attention mechanism. For instance, ``sparse attentions'' promote predefined sparse patterns such as sparse transformers \cite{child2019generating} and low-rank attentions use linear sketching such as linformers \cite{wang2020linformer}. Kernel methods also approximate attention with ensembles of kernels such as performers \cite{choromanski2020rethinking}.

Yet, these remedies fall short in handling the desired kilometer-scale resolution in weather and climate data. More recently, the Adaptive Fourier Neural Operator (AFNO) has provided a class of transformers that can handle continuous input and high resolutions efficiently \cite{guibas2021adaptive}. AFNOs possess the merits of neural operators and achieve superior scaling characteristics. FourCastNet employs the AFNO architecture to achieve unprecedented accuracy and resolution.

\subsubsection{Data-driven surrogates}

Recent studies have shown that DL can create data-driven surrogates of weather and climate models that are orders-of-magnitude computationally less expensive to run than traditional numerical models \citep{weyn2021sub,rasp2021data}. The advantages of this alternative approach include overcoming model biases inherent in NWP models and rapidly generating large ensembles due to high throughput via batched inference.

Until recently, however, the forecast accuracy and resolution of DL surrogates were far inferior to state-of-the-art operational weather and climate models. FourCastNet has closed the resolution disparity significantly and has demonstrated superior forecast accuracy compared to operational weather models at short lead times on certain atmospheric variables \cite{FCN}. FourCastNet has eight times greater resolution (25 km vs. 200 km) and is significantly more accurate than the most recent state-of-the-art DL global weather surrogate (cf. Figure~\ref{fig:dlwp}). Due to its high resolution, FourCastNet resolves extreme events such as tropical cyclones and atmospheric rivers and captures fine-scale phenomena that have been inadequately represented by prior DL models due to their coarser grids. 

\begin{figure}
    \centering
    \includegraphics[width=0.4\textwidth]{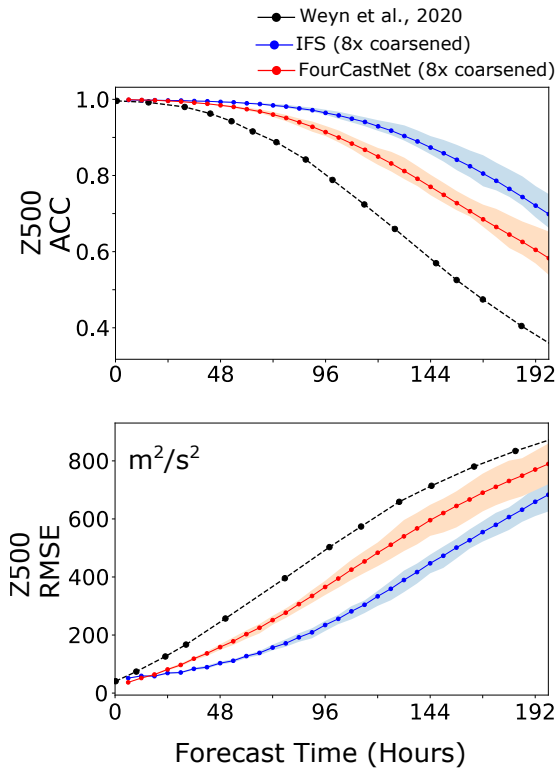}
    \caption{\small Comparison of performance metrics (ACC: Anomaly Correlation Coefficient and RMSE: Root Mean Squared Error) between downsampled FourCastNet predictions, downsampled IFS, and baseline state-of-the-art deep learning weather prediction (DLWP) model \citep{weyn2021sub} for $Z_{500}$, geopotential height at 500 hPa, a key determinant of global weather patterns. FourCastNet significantly outperforms DLWP and predicts at 8X higher resolution.}
    \label{fig:dlwp}
\end{figure}

At inference time, FourCastNet requires 12.41 seconds to generate a 100-member ensemble forecast on a single Selene node with eight NVIDIA A100 GPUs. This translates to a latency of 12.41 node-seconds for FourCastNet with parameters as described in Section~\ref{sec:innovations}. By comparison, the IFS L91 18 km model requires 984,000 node-seconds for a 100-member ensemble forecast~\cite{bauer2020ecmwf}. Thus, FourCastNet generates forecasts that are \emph{80,000} times faster than the IFS on a node-seconds basis. Comparing the power consumption of the respective chips used in the IFS and FourCastNet forecasts, we estimate that the IFS ensemble forecast consumes $2.71 \times 10^8$ J, whereas the FourCastNet ensemble forecast consumes $2.98\times 10^4$ J. Thus, FourCastNet is \emph{10,000} times more energy efficient than the IFS. Finally, once trained, FourCastNet requires only a single GPU node to generate ensemble forecasts in contrast to more than 1,000 nodes required by the IFS.

\section{Innovations}
\label{sec:innovations}

\subsection{Model Innovations}
\begin{figure}
    \centering
    \includegraphics[width=0.5\textwidth]{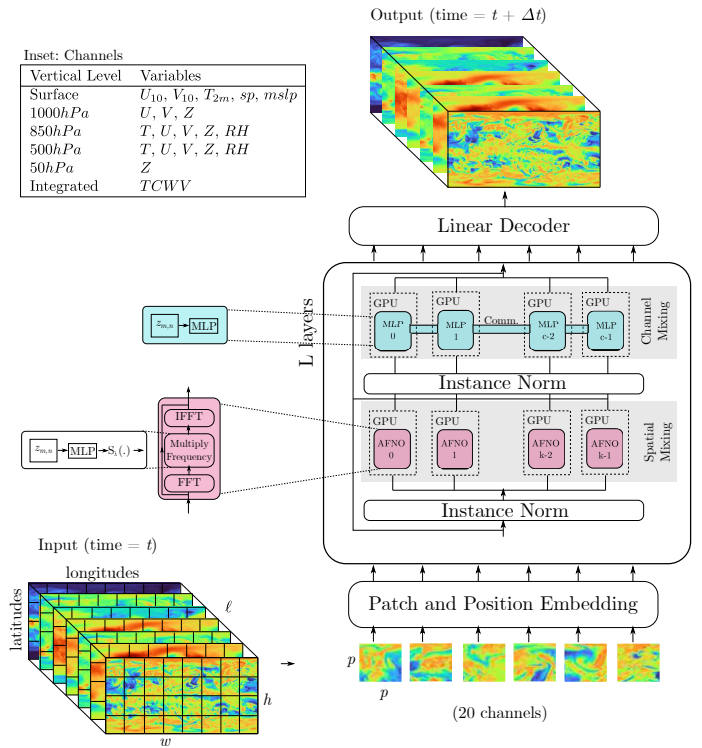}
    \caption{\small The AFNO architecture showing the key operations performed on the the input tensor with dimensions ($20 \times 720 \times 1440$) to produce a 6 hour single time step forecast with the same dimensions. Model parallelism is implemented by splitting the channels (feature maps) across GPUs. Channel mixing MLP operations require communication across the model parallel ranks, while the FFT based spatial-mixing operates on disjoint blocks that are embarrassingly parallel.} 
    \label{fig:afno}
\end{figure}

The AFNO transformer works as follows. The input frame $\vv{X}(t)$ with spatial resolution $h \times w \times \ell$ ($h=720$, $w=1440$, $\ell=20$) is first divided into a sequence of patches or tokens, each with $p\times p \times \ell$ pixels as shown in Figure~\ref{fig:afno}. Since vision transformers generally become more powerful with smaller patch sizes, we set $p=4$. This yields a grid of 180$\times$360 tokens. Each token is then embedded in a high dimensional vector (e.g., with 1,024 numerical entries) which is passed to a series of transformer layers to refine the embedding. Finally, the pixels for frame $\vv{X}(t+\Delta t)$ at time $t + \Delta t$ are reconstructed from the embedding of the last layer using a linear decoder. Each transformer layer also applies spatial (token) mixing and channel mixing. The channel mixer consists of a single-layer multi-layer perceptron (MLP). The key component of AFNO is a carefully designed spatial mixing operation that entails the following steps:

\vspace{1mm}

\begin{enumerate}
    \item An FFT spatially mixes the 180$\times$360 grid of tokens.
    \item For each individual token, the 1024 channels are mixed using a block-diagonal MLP in the Fourier space. All $64,800$ tokens are processed in parallel, where the MLP weights are shared across tokens.
    \item An inverse FFT demixes the tokens and returns them to the spatial domain.
\end{enumerate}

\subsection{Training and Inference}
We train the DL model on the ERA5 reanalysis dataset from the ECMWF at the native resolution of $0.25^\degree$ lat-lon on a regular cartesian grid. At a given timestep, each variable is represented as a $720 \times 1440$-pixel field. We sub-sample the available dataset to include the 20 prognostic variables thought to most strongly influence near-surface winds and temperatures. These variables are listed in the inset of Figure~\ref{fig:afno}. We sub-sample the data at 6-hour intervals from the set of 1-hour intervals available in ERA5. Denoting the modeled variables by the tensor $\vv{X}(t)$ with dimensions $20 \times 720 \times 1440$, we train the AFNO model to learn the mapping from $\vv{X}(t)$ to $\vv{X}(t+\Delta t)$ with a fixed $\Delta t$ = 6 hours. The 20 variables are treated as input channels for the AFNO model. Data from the years 1979--2017 are used for model training and hyper-parameter optimization, while the data from 2018 onwards is held out for final testing. In the pre-training phase, we train the model for 80 epochs using an $\ell_2$ loss. In the fine-tuning phase, we start from the previously pre-trained model and optimize it to predict two time steps. The model first generates the output $\vv{X}(t+1)$ from the input $\vv{X}(t)$. The model then uses its own output $\vv{X}(t+1)$ as input and generates the final output $\vv{X}(t+2)$. We then compute a training loss by comparing each of $\vv{X}(t+1)$ and $\vv{X}(t+2)$ to the respective ground truth from the training data and sum the two for model optimization.

\subsection{Implementation Details}
We implement FourCastNet in the PyTorch framework \cite{website:pytorch} which is a domain specific language for DL. PyTorch offers both rapid prototyping capabilities as well as low-level operations for performance optimization. We will discuss the most important implementation details in the following.

a) \emph{Parallelism}: We employ model parallelism in the form of feature parallelism; we split the feature maps/channels in most operations where this is possible/feasible. Operations which do not mix feature maps such as the FFTs become embarrassingly parallel, while those which do mix feature maps require additional communication (e.g. MLP for intra-patch mixing). The latter normally requires different function calls for forward and backward passes. For example, a row-parallel matrix multiplication does not require communication in the forward pass but it requires an output gradient reduction step in the backward pass. We use PyTorch's custom autograd functions to implement the model-parallel portions of this code including gather, scatter, and reduction routines. We refer to a collection of GPUs which share the same model weights as a model instance.
In addition to feature parallelism, we also apply data parallelism, splitting the global batch across independent model instances. In the backward pass, weight gradients need to be reduced across all model instances. 
One important aspect of feature parallelism is that weight gradients do not need to be reduced between GPUs in the same model instance. Instead, each model parallel rank $k$ needs to be coupled with the matching rank in the other instances. We exploit this fact by implementing orthogonal communicators for data and model parallelism. Model parallel communications require lower latencies and higher bandwidth, because they mostly cannot be hidden behind computation. Therefore, all GPUs from the same model instance reside on the same node to exploit the high-bandwidth, low-latency NVLink interconnect between them. The communication backend of PyTorch employs the highly-optimized NVIDIA Collective Communication Library (NCCL). NCCL provides abstractions for Infiniband and NVLINK alike, as well as hybrid algorithms leveraging both interconnects simultaneously in order to improve the effective bandwidth.

b) \emph{Spectral convolution}: The spectral convolution is an integral part of the network architecture and PyTorch does not yet provide a specialized layer for it. Therefore, we implement this layer using real-to-complex and complex-to-real FFT layers for the spectral forward and backward transformation. Note that the FFTs are batched over both batch and feature dimensions. 
The MLP in spectral space is implemented using the \texttt{torch.einsum} function call, which allows the user to specify arbitrary tensor index contractions via Einstein summation index conventions. The cost of this kernel can vary by up to 400\%, depending on the data layout of the weight and input tensors. This is because PyTorch translates the \texttt{torch.einsum} function call into a series of data layout transformations and batched matrix multiplications.
Since the weight tensors as well as inputs are complex valued, we found that it is most efficient to use a memory layout which allows PyTorch to generate a single batched complex GEMM call from cuBLAS. While it is currently not possible to use complex weight tensors natively in PyTorch due to implementation limitations of the weight update step, it is straightforward to store the weights as real tensors with real and imaginary components as the fastest index. The helper functions \texttt{torch.view\_as\_complex} and \texttt{torch.view\_as\_real} allow switching between real valued and complex valued views without any data copy overhead. We apply the same trick to the activation functions, which then act on real and imaginary parts independently.
We apply feature parallelization to this layer which effectively splits the batch of the FFTs across multiple GPUs. The same happens to the spectral convolution step: since our degree of model parallelism ($<=8$, see below) is smaller than the number of blocks in the block-diagonal MLP (32), we can distribute the blocks across the GPUs. These two properties render this layer embarrassingly parallel in feature parallelization mode.

c) \emph{MLP}: Each AFNO transformer block uses an MLP to correlate patch information locally and across heads. Using the NCHW data layout, we implement the matrix multiplications in the MLP using pointwise convolution layers (i.e. convolutions with unit stride and kernel size). 
Since the hidden dimension with 4,096 units is four times bigger than the embedding size with 1,024 units, we employ row-wise parallelism to the first matrix multiplication and column-wise parallelism to the second. 

d) \emph{I/O pipeline}: The preprocessed dataset is comprised of a collection of HDF5 files where each file contains a row-major array holding one year of data. The data needs to be loaded and preprocessed by normalizing with the mean and variance per-variable and then augmenting with additive gaussian noise.  
We employ the external source functionality of NVIDIA DALI \cite{website:nvidia-dali} to implement this pipeline in an efficient fashion. Inside the external source data loader, we use double buffering with pinned host buffers to enable DALI to copy data directly to the GPU via RDMA without intermediate CPU buffer copies. For pinning the host buffers directly from python, we use the NVIDIA CuPy package.
The remaining steps of the preprocessing pipeline run concurrently to the training step on the GPU. For our application, it is sufficient to use a single external source process and assign two threads to the data pipeline.

e) \emph{CUDA graphs}: Aside from I/O, the biggest source of performance variability at scale originates from CPU interference, e.g. operating system context switches delaying kernel launches and thus stalling the GPU. To minimize these effects, we capture the full forward and backward step of the training process inside a CUDA graph. This records the full sequence of kernel launches and thus prevents the CPU from interfering with kernel launches. The remaining code susceptible to this interference is the I/O pipeline, the optimizer step, and validation. The I/O pipeline runs concurrent to the training step and is much faster, thus a CPU interference is negligible here. The optimizer only comprises 5-8\% of the overall training time and thus we refrained from capturing it. Lastly, validation is embarrassingly parallel across data parallel groups and thus stalls only affect the respective model instance.

f) \emph{JIT compilation}: Some composite layers and metrics, such as the ACC metric and the RMSE loss, involve performing a series of lightweight mathematical operations, such as multiplying values with or adding values to a tensor. Each of these operations generates a separate kernel call in PyTorch. We use \texttt{torch.jit.script} to fuse blocks of these operations into larger kernels to reduce launch latency and improve memory locality, especially for the validation section of the code which is not captured in a CUDA graph.

\section{Performance Measurement}
\label{sec:performance_measurement}

\subsection{Measurement Methodology}
In most DL training applications, the relevant performance metric is \emph{time-to-solution}: the time it takes to train the network. In DL, training is considered complete once the neural network reaches a desired score in some scoring metric. This could be a certain loss value, target score of an accuracy metric, or more general objective function. In our case, we are interested in comparing pre-training performance, as pre-training constitutes a substantial fraction of the total compute cost of training FourCastNet from scratch. Pre-training is defined as training the network over a fixed number of epochs to do single-step prediction, so that it can be later refined in a more expensive multi-timestep training procedure. We found that \PretrainingEpochs epochs are a good target. This number was chosen empirically to optimize forecasting skill at 192-hour lead time, after fine-tuning.

Unfortunately, it is computationally prohibitive to train the network for \PretrainingEpochs epochs at all scales on all systems, so we strategically choose training runs as follows. First, we determine a set of hyper parameters (e.g. batch sizes, learning rates) which lead to a good pre-trained model configuration with comparable generalization errors after \PretrainingEpochs training epochs. Then, we use those hyper parameters to scale out our pre-training runs, either by increasing batch size and thus increasing the degree of data parallelism, or by increasing the degree of model parallelism. We then train the network at each of these scales for \ScalingEpochs epochs in order to obtain per epoch run time statistics. Since the number of epochs to convergence is the same for all scales, the per epoch time is directly correlated with overall training time.

Even at large scale, epoch times span tens of seconds to a few minutes, and therefore the python \texttt{time.time} function is sufficient to measure timings. Our timers measure epoch time (including validation) and the overall training time, including exposed IO overhead. 

We nevertheless exclude scaffolding time from the measurements. This means, our timers do not account for time spent in:
\begin{itemize}
    \item wireup/bootstrapping of the NCCL communicators
    \item warming up the network for 20 steps for cuDNN auto-tuning and JIT compilation of selected kernels and metrics
    \item performing the CUDA graph capture of forward and backward pass
\end{itemize}
In practice, we find that the scaffolding time is negligible compared to training time.

We use NVIDIA Nsight Compute to measure floating point operations. More precisely, we use a custom \textit{ncu} section file that specifies {\scriptsize \texttt{sass\_\_inst\_executed\_per\_opcode\_with\_modifier\_all}} as the metric to be used in order to obtain a list of all the instructions executed within each kernel run on the GPU. We then count the FLOPS generating instructions and multiply them by their specific weighting factors. The weighting factor for a multiplication or addition is 1, the weight for FP32 and FP64 fused multiply-and-add (FMA) is 2 and the weight for an FP16 double FMA is 4. For the tensor core (xMMA type) instructions, the weight is computed as $2{\times} M {\times} N{\times} K$, where M, N, K are the dimensions of the contracted tensors. Thus for FP16 and TF32 type tensor operations (HMMA) we have $2{\times} 16{\times} 8{\times} 8 = 2,048$ and $2 {{\times} 16{\times} 8{\times} 16} = 4,096$ FLOPS respectively. We do not count integer operations. The FLOPS weight factors are summarized in Table \ref{tab:flop_weights}.

\begin{table}[ht]
    \centering
    \begin{tabular}{c|c|c}
    Precision & Instruction & Weight Factor \\\hline
    \multirow{3}*{FP64} & \texttt{DADD} & 1 \\
    & \texttt{DMUL} & 1 \\
    & \texttt{DFMA} & 2 \\\hline
    \multirow{3}*{FP32} & \texttt{FADD(.FTZ)} & 1 \\
    & \texttt{FMUL(.FTZ)} & 1 \\
    & \texttt{FFMA(.FTZ)} & 2 \\\hline
    \multirow{2}*{FP16} & \texttt{HADD2.F32} & 1 \\
    & \texttt{HFMA2.MMA} & 4 \\\hline
    \multirow{2}*{Tensor Core FP16} & \texttt{HMMA.1688.F32} & 2048 \\
    & \texttt{HMMA.16816.F32} & 4096 \\\hline
    Tensor Core TF32 & \texttt{HMMA.1688.F32.TF32} & 2048 \\
    \end{tabular}
    \caption{FLOPS weights for various GPU instructions relevant to our code.}
    \label{tab:flop_weights}
\end{table}

In order to obtain the overall FLOPS count for a single iteration, we multiply the single GPU FLOPS count by the total number of GPUs. This procedure is accurate as the training is perfectly load-balanced across all model and data parallel ranks. We then multiply the per-iteration FLOPS count by the number of performed iterations to obtain the total FLOPS counts for the entire training. We also account for validation FLOPS using the same approach.
We compute a FLOPS rate per epoch by adding the training and validation FLOPS counts and dividing by the total epoch duration, i.e. training and validation time. For each run, we compute the median and max FLOPS rate over epochs and define this to be our sustained and peak FLOPS rates respectively.  
Note that this underestimates the peak FLOPS rate, since it is averaged over a full epoch. Furthermore, we do not count additional FLOPS attributed to the weight gradient reductions across data parallel groups as well as any FLOPS generated by the CPUs. Both of these values are just a small fraction of the overall FLOPS count.

\subsection{Systems and Environment}\label{sec:systems_and_environment}

\subsubsection{Perlmutter}
Perlmutter \cite{website:perlmutter} is a hybrid supercomputer based on the HPE Cray Shasta platform installed at Lawrence Berkeley National Laboratory. At time of this submission, only Phase-1 of the staged deployment was finished and we only use the GPU partition for our experiments. 
This partition consists of 1,536 nodes, each equipped with 4 NVIDIA A100-40 GB GPUs connected via NVLink3 with an all-to-all topology. Furthermore, each node has one AMD EPYC 7763 CPU, featuring 128 logical cores at 2.45 GHz as well as 256 GB DDR4 DRAM. Phase-1 Perlmutter uses the HPE Cray Slingshot 10 interconnect with 2 HCA per node.
There is no node-local storage available but Perlmutter uses an all-flash, LUSTRE-based global file system for data intensive tasks. As of November 2021, Perlmutter is ranked fifth on top500, quoting 93.8 petaFLOPS linpack peak performance.

\subsubsection{JUWELS Booster Module}
JUWELS Booster Module \cite{JUWELS} is an Atos BullSequana XH2000 system installed at Jülich Supercomputing Center. It is comprised of 936 compute nodes, each equipped with 4 NVIDIA A100-40 GB GPUs. Similar to Perlmutter, the GPUs are connected via NVLink3 in an all-to-all fashion. Each node is dual socket featuring two AMD EPYC 7402 with 48 logical cores each at 2.8 GHz and has a total of 512 GB DDR4 RAM (i.e. 256 GB per socket). Each node has 4 NVIDIA HDR200 InfiniBand ConnectX-6 HCA, connecting the system in a DragonFly+ topology.
The global file system is GPFS based with a peak bandwidth of ~1 TB/s. Additionally, the High-performance Storage Tier (HPST) supports an additional NVMe-based cache layer based on the DDN Infinite Memory Engine (IME) technology. This layer is offering high bandwidth and low latency access to pre-staged data, which we make use of for our experiments. As of November 2021, JUWELS Booster Module is ranked eighth on top500, quoting 71.0 petaFLOPS linpack peak performance.

\subsubsection{Selene}
Selene is built from 4 DGX SuperPODs \cite{website:superpod} and installed at NVIDIA headquarters in Santa Clara. It is comprised of 560 DGX-A100 nodes each featuring 8 NVIDIA A100-80 GB GPUs. The GPUs are partitioned into two groups of 4. Within a group, all GPUs are all-to-all connected with NVLink3, similar to the previously mentioned systems. In addition, the edges of both groups are also connected with NVLink3. 
The nodes are dual socket AMD EPYC 7742 with 128 logical cores each at 2.25 GHz and a total of 2 TB DDR4 memory. The system has 8 NVIDIA HDR200 InfiniBand ConnectX-6 HCA per node and is connected in a full fat-tree topology. Each node is equipped with 32 TB node local NVME storage. The global LUSTRE file system offers a peak bandwidth of about 1 TB/s.
As of November 2021, Selene is ranked sixth in top500, quoting 79.2 petaFLOPS linpack peak performance. 

\subsubsection{Software environment}
All our experiments use the NVIDIA NGC PyTorch base container in version 22.02 \cite{website:ngc-pytorch}. 
The individual software versions of relevant packages are PyTorch v.1.11, CUDA v11.6.55, cuDNN v8.3.2, cuFFT v10.7.0, cuBLAS v11.8.1, NCCL v2.11.4, CuPY 9.6.0 and Python 3.8.12. We further installed a nightly version of DALI tagged 1.13.0dev.20220324.
We run the docker container directly on Selene using \texttt{pyxis/enroot} and on Perlmutter using \texttt{Shifter}. For JUWELS, we convert the docker container into a singularity image before running.

\section{Performance Results}
\label{sec:performance_results}

\subsection{Performance: Single Model Instance}
We collect the FLOPS counts for a single iteration and a single GPU for all model instance sizes using batch size 1. An iteration includes data loading and preprocessing, forward and backward pass of the neural network, the weight update step, and a learning rate scheduler step. 
A breakdown of FLOPS per GPU and iteration, precision and model instance size can be found in Table \ref{tab:flop_breakdown}. For the sake of brevity, we have combined all FLOPS from tensor instruction into a single line item \emph{tensor}.

\begin{table}[ht]
    \centering
    \begin{tabular}{c|c|c|c|c}
       Model instance size & Phase & FP32 & FP16 & Tensor \\\hline
         \multirow{2}*{1} & training & 28 & 6 & 8298 \\
         & validation & 13 & 4 & 2840 \\\hline
         \multirow{2}*{2} & training & 14 & 3 & 4160 \\
         & validation & 7 & 2 & 1424 \\\hline
         \multirow{2}*{4} & training & 8 & 2 & 2088 \\
         & validation & 4 & 2 & 670 \\\hline
         \multirow{2}*{8} & training & 4 & 2 & 1036 \\
         & validation & 2 & 1 & 339 \\
    \end{tabular}
    \caption{FLOPS count per iteration per GPU for different model instance sizes and precisions in units of $10^9$.}
    \label{tab:flop_breakdown}
\end{table}

Most FP32 and TF32 FLOPS are generated by the FFT layers, while most of the FP16 FLOPS are generated by the dense matrix multiplications in the MLP layers. Although we account for FP64 FLOPS in our total FLOPS count, they are not shown in this table because their share is negligible. The source of these FLOPS can be traced to the optimizer weight update step. As expected, the FLOPS count per GPU decreases by roughly the same factor as the degree of model parallelism increases. Furthermore, the table shows that the majority of FLOPS are executed on tensor cores, which indicates that most operations can take advantage of the special acceleration logic.

Since a single epoch can take very long for these small scale configurations, we measured training step timings for about 100 steps averaged over blocks of 10 steps each.

On a single GPU (model instance size 1), the average step time is about 517 ms, translating into a performance of 160 TFLOP/s, which is about 51\% of dense FP16 tensor core peak performance (quoted with 312 TFLOP/s in the A100 whitepaper). 
For model instance size 2, the performance per GPU drops slightly to 115 TFLOP/s and then to 76 TFLOP/s and for model instance sizes 4 and 8 respectively. This performance drop is mostly due to the additional communication in forward and backward passes, which cannot be overlapped with other computations since our neural network is strictly feed-forward and does not feature branches with significant amounts of compute. A detailed breakdown of operation types where time is spent for a training step is reported in Table 
\ref{tab:time_breakdown}

\begin{table*}[ht]
    \centering
    \begin{tabular}{c|c|c|c|c}
    Model instance size & Conv/Matmul & FFT & Communication & Elementwise Operations \\\hline
    1 & 37 & 9 & 0 & 42 \\
    2 & 29 & 7 & 19 & 35 \\
    4 & 20 & 4 & 34 & 32 \\
    8 & 14 & 3 & 51 & 24 \\
    \end{tabular}
    \caption{Time breakdown in \% for most important operations for different model instance sizes.}
    \label{tab:time_breakdown}
\end{table*}

\begin{figure}
    \centering
    \includegraphics[width=0.4\textwidth]{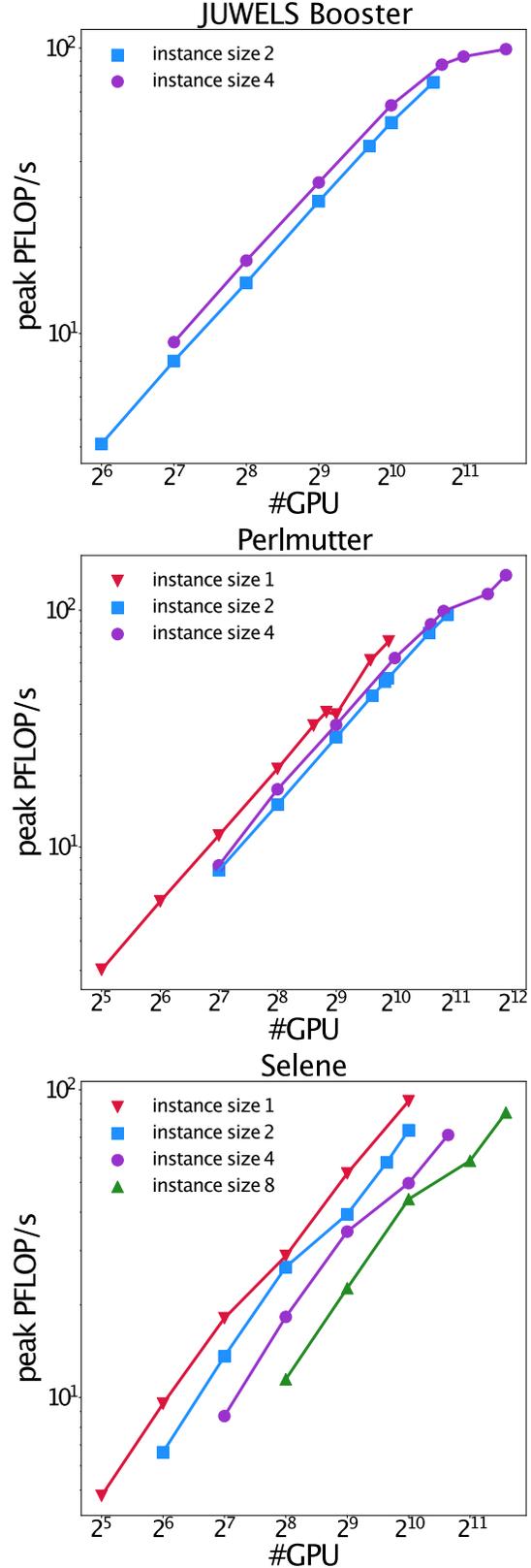}
    \caption{\small FourCastNet scaling on JUWELS Booster (top), Perlmutter (center) and Selene (bottom) for various model instance sizes.}
    \label{fig:fourcastnet_scaling}
\end{figure}

The communication column also includes kernels necessary for postprocessing communicated data, for example the concatenation time for all-gathered tensors. The table shows that communication becomes the dominant part of the iteration step for bigger model instance sizes. FFT time is always subdominant because the AFNO patch embedding step generates grids with very small FFT dimensions.

\subsection{Scalability}

We perform several scaling experiments on all three systems described in section \ref{sec:systems_and_environment}.
We always bind the GPUs to the closest available cores, memory and interconnects. On Selene, we additionally ensure that the model instances are partitioned so that GPUs on the same socket ideally belong to the same model instance. This is only relevant for model instance sizes 2 and 4, since groups of size 8 span full Selene nodes. On Perlmutter and JUWELS Booster, all GPUs are all-to-all connected and thus we do not need any additional partitioning configuration. We only run scaling experiments where the model instance spans at most a full node. On JUWELS we found that atomic model instances do not run due to memory constraints, whereas on Perlmutter this was not a problem. Some CUDA contexts are created by default on JUWELS Booster GPUs which drive the total allocated memory above the maximum capacity.

Figure \ref{fig:fourcastnet_scaling} shows the results of our scaling experiments. We observe that model instances of size 4 perform better than those of size 2 on JUWELS Booster and Perlmutter. On Selene, the picture is different: the performance gain is bigger when data parallelism is increased compared to model parallelism. This is due to the network interconnect design: Selene has balanced intranode and internode bandwidth. This allows NCCL to use hybrid NVLink/Infiniband algorithms which are able to effectively use a large percentage of the intranode NVLink bandwidth at large scales. Therefore, Selene does not suffer as much from large scale collective operations as Perlmutter and JUWELS Booster do. On the latter systems, it is therefore preferable to introduce more local communications than to increase the global communication scale further.
In the Selene plots one can observe ``kinks'' in the scaling curves at 128, 256, 512 and 1,024 GPU points for model instance sizes 1, 2, 4 and 8 respectively. This can be explained by NCCL switching from ring- to tree-based reduction algorithms at 128 GPU data parallel scale, \emph{i.e.} for model instances of size $n$ this switch occurs at $n{\times} 128$ total GPU scale. 
On Perlmutter, the data point for model instance size 1 on 512 GPU is lower than expected probably due to interference with other jobs running concurrently on the system. We observe that scaling on JUWELS Booster tapers off beyond 1,024 GPUs whereas Selene and Perlmutter continue to scale. At 1,024 GPUs, JUWELS Booster is about 13\% slower than Perlmutter but this gap increases to 18\% at 3,072 GPUs.
The wireup and scaffolding time grows linearly with number of GPUs; on Perlmutter at 1,568 GPU scale, the initialization takes about 4 minutes and at maximum scale, i.e. 3,808 GPUs, it takes about 11 minutes total.

The peak performances measured are \PeakFLOPJUWELS PFLOP/s on \PeakFLOPJUWELSScale GPUs, \PeakFLOPPerlmutter PFLOP/s on \PeakFLOPPerlmutterScale GPUs, and \PeakFLOPSelene PFLOP/s on \PeakFLOPSeleneScale GPUs for JUWELS Booster, Perlmutter, and Selene respectively.

\subsection{Convergence and Time-to-Solution}

\begin{figure}
    \centering
    \includegraphics[width=0.41\textwidth]{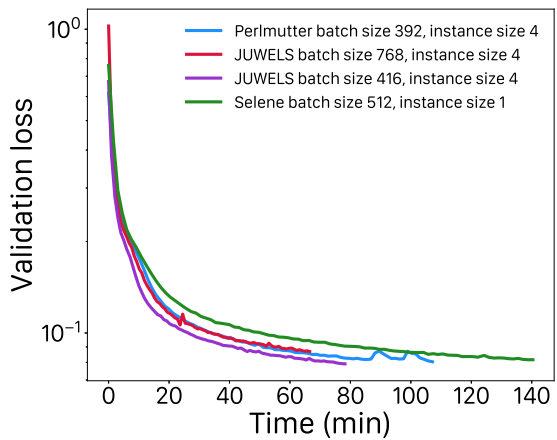}
    \caption{\small Validation loss as a function of wall-clock time for various FourCastNet configurations on JUWELS Booster, Perlmutter and Selene. The plot shows a significant reduction in solution times as parallelism increases.}
    \label{fig:fourcastnet_convergence}
\end{figure}

The efficient scaling properties of FourCastNet allows training larger models more quickly by consuming more resources for shorter amounts of time. We estimate the pre-training wall-clock time for a baseline configuration using no model parallelism and a batch size of 32 to be roughly 40 hours. By comparison, Figure~\ref{fig:fourcastnet_convergence} shows convergence curves of the validation loss as a function of wall-clock time for various configurations on JUWELS Booster, Perlmutter, and Selene. Clearly, we can substantially reduce training time-to-solution to just over an hour by increasing the degree of model parallelism without incurring much loss of efficiency. We note that hyper-parameters for batch size 768 are based on extrapolations estimated from hyper-parameter optimizations up to batch size 512. While we observe excellent convergence at batch size 768, this could likely be improved with further tuning at this scale. The dramatic reduction in time-to-solution gained by running at scale enables much more rapid exploration of hyper-parameters and model experimentation for practitioners.

\section{Implications}
\label{sec:implications}

\subsection{Implications for Weather and Climate Science}

FourCastNet has achieved five orders-of-magnitude speedup over traditional numerical simulations while maintaining high accuracy, with significant implications for weather and climate science, future computing systems, and society. 

\emph{Complexity}: FourCastNet shows conclusively that complex multi-scale, multi-physics turbulent spatio-temporal dynamics of global weather systems can be emulated using state-of-the-art DL with the principled physics-inspired approach of FNOs. A whole Earth system emulator that can mimic other components such as ocean, sea ice, and land surface is not far in the future.

\emph{Resolution, dimensionality, ensemble size:} FourCastNet's step-change in resolution and dimensionality implies that emulating the Earth system, resolving fine-scale processes, and predicting impacts at scales that matter are within reach. The scalability of FourCastNet will be key for emulating data at 1 km-scale, the target for exa-scale weather and climate computing. Indeed, the true power of the AFNO model, and transformers in general, is realized at scale when (i) the model has hundreds of billions of parameters, larger embedding dimensions, deeper networks, and smaller patch sizes \cite{riquelme2021scaling, khan2021transformers}; and (ii) the training dataset is massive. As the resolution grows, the patch size of the AFNO can be reduced from $p=4$ to $p=1$ (the one-pixel resolution) to capture fine-scale behavior accurately and model uncertainties. Thus, although FourCastNet today emulates weather at 25 km, the success of the hybrid data and model parallelism framework shown in this work and FourCastNet's scalability to larger batch sizes, smaller patch sizes, and more GPUs makes it well-positioned for the extreme-scale training needed to address the grand challenge of kilometer-scale emulation. Furthermore, the orders-of-magnitude speedup in inference implies that large ensembles can be generated in seconds and that the tera-dimensional phase space can be explored in near real time. This will enable confident prediction of extreme weather and climate events that cannot be reliably predicted in smaller ensembles. 

\emph{Throughput and the data avalanche:} FourCastNet's time-to-solution, the category of achievement that this work is nominated for, is a breakthrough for the enterprise of Earth system modeling because it blasts through the biggest bottlenecks in traditional numerical simulations: throughput and the data avalanche. Although training FourCastNet at kilometer-scale will be expensive, because inference is cheap, predictions can be made at close to 100,000 SYPD at the same computational cost as 1 SYPD from traditional numerical simulations. A single century-long trajectory of a km-scale climate simulation with outputs written every 30 simulated minutes will generate an exabyte of data. Schar et al. \cite{Schar2020} emphasize that data movement and storage constraints, the so-called data avalanche, will be the grandest challenge in the era of exascale weather and climate computing. They propose storing the simulation setup, initial conditions and restart files (checkpoints) to rerun simulations on demand. FourCastNet's throughput makes it perfectly poised to reduce the data avalanche down to a trickle by enabling on-demand near real-time predictions from checkpoints at breakneck speed. As the limits of predictability of data-driven models like FourCastNet increase from weather to climate timescales, even fewer checkpoints will be needed to ``tether'' the emulator to climate trajectories.   

\emph{Extrapolation:} The Earth system is non-stationary and Earth's climate is changing. The quality and density of Earth observations degrade rapidly moving backwards through time. For reasons described in Section~\ref{sec:overview}, generating sufficient high-fidelity, high-resolution simulations that span the extremely high-dimensional space of possible futures is nearly impossible. Data-driven DL emulators will need to use physical laws to extrapolate into unprecedented futures. The next frontier of data-driven DL models is physics-informed DL and there has already been rapid progress in this sub-field of scientific machine learning \cite{PINO}, including in applications to weather and climate modeling \cite{kashinath2021physics}. The next generation of FourCastNet will leverage physical laws to address the extrapolation challenge. 

\subsection{Implications for Future Supercomputing Systems}

Simulation with in-situ DL training and in-situ analytics are potential solutions for a host of challenges in the era of exascale weather and climate computing. FourCastNet makes this  a near-term possibility. Indeed, NVIDIA is already developing an AI supercomputer, Earth-2, to build digital twins of Earth. Powered by Earth-2 and driven by models such as FourCastNet, digital twins will predict regional climate change in real time and on demand, and enable the discovery of mitigation and adaptation strategies. Supercomputing systems like Earth-2 that need to simultaneously support high throughput of traditional numerical simulations, rapid streaming access to large observational datasets, fast data processing, and fast training of DL models require: (i) large on-node GPU memory and high-bandwidth low-latency interconnect; (ii) high external network bandwidth; (iii) reduced I/O bottlenecks and rapid data movement; (iv) optimized mixed precision including FP8; (v) a tiered architecture with transparent caches; and (vi) tight interaction between GPU and CPU. Spectral models such as IFS and the FNO will benefit from hardware that supports fast global all-to-all communication. Perhaps most importantly, such a supercomputer will require hardware-software co-design and rethinking the entire stack to provide dense integration of simulation, DL training, inference, and virtualization in platforms such as NVIDIA Omniverse.

\subsection{Implications for Society}

FourCastNet opens the door to \emph{Interactivity at Scale} with profound implications for how weather and climate data can enable well-informed action and empower not just scientists and experts but also society, including policy- and other decision-makers. Schar et al. \cite{Schar2020} describe the crucial importance of a virtualization layer to deal with the impending data avalanche that will bring exabytes of data that cannot be stored, nor interacted with, efficiently. In the era of exascale computing and exabytes of data, Bauer et al. \cite{Bauer2021} describe how a breakthrough is needed in today's information systems for information access and intervention, the core ingredients to build digital twins of Earth. FourCastNet holds tremendous promise for this breakthrough and represents a leap towards achieving virtualization and building digital twins of Earth that will allow users to intervene, extract information, and influence planet Earth's future, which is precisely what is needed to further climate action and empower individuals.

\section{Acknowledgments}
This research used resources from the National Energy Research Scientific Computing Center (NERSC), a U.S. Department of Energy Office of Science User Facility located at Lawrence Berkeley National Laboratory, operated under Contract No. DE-AC02-05CH11231. 
The authors also gratefully acknowledge the Gauss Centre for Supercomputing e.V. (www.gauss-centre.eu) for funding this project by providing computing time on the GCS Supercomputer JUWELS \cite{JUWELS} at Jülich Supercomputing Centre (JSC). We thank Peter Dueben (ECMWF), Peter Bauer (ECMWF), Bjorn Stevens (MPI-M), Pedram Hassanzadeh (Rice U.), Ashesh Chattopadhyay (Rice U.), and Torsten Hoefler (ETHZ) for many valuable discussions that have shaped this research. We are grateful for the support of staff at the Jülich Supercomputing Centre, NERSC, and the NVIDIA Selene team for their assistance with the runs on their supercomputing systems.


\bibliographystyle{IEEEtran}
\bibliography{IEEEabrv}

\end{document}